\documentclass{v16nufact}
\usepackage[latin9]{inputenc}
\usepackage{units}
\usepackage{amsmath}
\usepackage{graphicx}

\makeatletter

\def\@aabuffer{}
\def\author #1{\expandafter\def\expandafter\@aabuffer\expandafter
{\@aabuffer \small\rm      #1\relax \par}}
\def\address#1{\expandafter\def\expandafter\@aabuffer\expandafter
{\@aabuffer \small\it #1\relax 
\\
\Photo
\par\vspace{1em}}}

\def\maketitle{
\begin{center}
   {\bf \@title \par}       
   \vskip 2em                      
   \@aabuffer\relax
\end{center} \par
\gdef\@aabuffer{}
}

\def\abstracts#1{
\begin{center}
{\begin{minipage}{5.2truein}
                 \footnotesize
                 \parindent=0pt #1\par
                 \end{minipage}}\end{center}
                 \vskip 2em \par}

\fussy
\RequirePackage[a4paper,margin=2.5cm]{geometry}
\RequirePackage[english]{babel}
\RequirePackage{graphicx,url}
\RequirePackage[colorlinks=true,urlcolor=blue,linkcolor=black,citecolor=black]{hyperref}
\def\section{\@startsection {section}{1}{\z@}{-3.5ex plus -1ex minus 
    -.2ex}{2.3ex plus .2ex}{\bf }}
\def\subsection{\@startsection{subsection}{2}{\z@}{-3.25ex plus -1ex minus 
   -.2ex}{1.5ex plus .2ex}{\it }}

\def\@makefnmark{{$\!^{\@thefnmark}$}}

\renewenvironment{thebibliography}[1]
        {\begin{list}{\arabic{enumi}.}
        {\usecounter{enumi}\setlength{\parsep}{0pt}
         \setlength{\itemsep}{0pt} 
         \settowidth
        {\labelwidth}{#1.}\sloppy}}{\end{list}}

\def\@citex[#1]#2{\if@filesw\immediate\write\@auxout
        {\string\citation{#2}}\fi
\def\@citea{}\@cite{\@for\@citeb:=#2\do
        {\@citea\def\@citea{,}\@ifundefined
        {b@\@citeb}{{\bf ?}\@warning
        {Citation `\@citeb' on page \thepage \space undefined}}
        {\csname b@\@citeb\endcsname}}}{#1}}

\newif\if@cghi
\def\cite{\@cghitrue\@ifnextchar [{\@tempswatrue
        \@citex}{\@tempswafalse\@citex[]}}
\def\citelow{\@cghifalse\@ifnextchar [{\@tempswatrue
        \@citex}{\@tempswafalse\@citex[]}}
\def\@cite#1#2{{$\!^{#1}$\if@tempswa\typeout
        {IJCGA warning: optional citation argument 
        ignored: `#2'} \fi}}

\def\baselinestretch{1.0}
\ifx\selectfont\undefined
\@normalsize\else\let\glb@currsize=\relax\selectfont
\fi

\ifx\selectfont\undefined
\def\@singlespacing{%
\def\baselinestretch{1}\ifx\@currsize\normalsize\@normalsize\else\@currsize\fi%
}
\else
\def\@singlespacing{\def\baselinestretch{1}\let\glb@currsize=\relax\selectfont}
\fi

\long\def\@caption#1[#2]#3{\par\addcontentsline{\csname
  ext@#1\endcsname}{#1}{\protect\numberline{\csname
  the#1\endcsname}{\ignorespaces #2}}\begingroup
    \@parboxrestore
    \footnotesize
    \expandafter\let\expandafter\@tempa\csname @make#1caption\endcsname
    \ifx\@tempa\relax\let\@tempa\@makecaption\fi
    \@tempa{\csname fnum@#1\endcsname}{\ignorespaces #3}\par
  \endgroup}
\long\def\@makefigurecaption#1#2{%
 \vskip 10pt 
 \setbox\@tempboxa\hbox{#1 -- {\footnotesize #2}}%
 \ifdim \wd\@tempboxa >\hsize #1 -- {\footnotesize #2}\par \else
  \hbox to\hsize{\hfil\box\@tempboxa\hfil} %
  \fi
 }
\newcommand{\lyxaddress}[1]{
\par {\raggedright #1
\vspace{1.4em}
\noindent\par}
}

\usepackage{units}

\def\Journal#1#2#3#4{{\em #1} {\bf #2}, #3 (#4)}

\def\be{\begin{equation}}
\def\ee{\end{equation}}
\def\bea{\begin{eqnarray}}
\def\eea{\end{eqnarray}}

\@ifundefined{showcaptionsetup}{}{%
 \PassOptionsToPackage{caption=false}{subfig}}
\usepackage{subfig}
\makeatother

\begin{document}
\vspace*{4cm}

\title{Recent Cross Section Work from NOvA}

\author{J. Wolcott, for the NOvA Collaboration}
\maketitle

\lyxaddress{\begin{center}
Department of Physics and Astronomy, Tufts University \\574 Boston
Ave., Medford, MA 02155 USA
\par\end{center}}

\abstracts{The NOvA experiment is an off-axis long-baseline neutrino
oscillation experiment seeking to measure $\nu_{\mu}$ disappearance
and $\nu_{e}$ appearance in a $\nu_{\mu}$ beam originating at Fermilab.
In addition to measuring the unoscillated neutrino spectra for the
purposes of predicting the oscillated neutrino spectrum in the far
detector, the 293-ton near detector also enables high-statistics investigation
into neutrino scattering in numerous reaction channels. We discuss
the various near detector analyses currently in progress, including
inclusive measurements of both electron and muon neutrino charged-current
interactions and efforts to constrain the off-axis NuMI flux using
the elastic scattering of neutrinos from atomic electrons.}

\section{Introduction}

Over the course of the last three decades, neutrino oscilliation experiments
have sought to use the quantum-mechanical properties of the neutrino
as a probe of the fundamental nature of the lepton family. Since the
weak-force coupling of neutrinos to other particles is extremely small,
terrestrial neutrino oscillation experiments, such as NOvA, typically
construct large detectors from materials composed of heavy nuclei
in an effort to maximize the neutrino interaction rate. But the intractibility
of calculating the dynamics of nucleons within the nucleus in the
low-energy limit of the strong force introduces significant uncertainties
into the reaction predictions used in measurements made with these
detectors. Even in the two-detector paradigm used by NOvA and other
experiments, in which a detector close to the neutrino source (the
near detector, ND) is used to constrain the product of interaction
cross section models and the flux prediction (which is then extrapolated
to the far detector, FD, where oscillations are observed), direct
measurements of neutrino interaction cross sections on the target
materials are extremely valuable for constraining and choosing between
models.

The 293-ton NOvA near detector is an ideal instrument to use for this
sort of cross section measurement for several reasons. First, its
location \unit[14.6]{mrad} off-axis in the Fermilab NuMI neutrino
beam it samples yields a narrow neutrino energy spectrum centered
on \unit[2]{GeV}, producing an event sample rich in interaction types
(including copious examples of quasielastic scattering, baryon resonance
production, and deep inelastic scattering) and exhibiting multiple
kinds of nuclear effects (including coherent meson production, multi-nucleon
scattering, and final-state hadron rescattering). Second, the detector
itself is a mostly-active, fine-grained, segmented tracking calorimeter
constructed of PVC cells filled with liquid scintillator with excellent
spatial and energy resolution. We present status reports on a number
of measurements currently in progress using the NOvA ND.

\section{$\nu_{\mu}$ charged-current inclusive scattering}

During its lifetime the NOvA ND is expected to record an immense sample
of charged-current (CC) interactions of muon neutrinos on the liquid
scintillator ($\nu_{\mu}CH_{2}\rightarrow\mu^{-}X$) ultimately numbering
in the millions. The statistical power of this sample offers an unprecedented
opportunity both to verify the basic nucleon-level models for CC reactions
in detail and to examine the relevant nuclear effects near $E_{\nu}=\unit[2]{GeV}$;
this energy range has previously been explored mostly in light bubble
chamber experiments in measurements reporting only total cross sections.

\subsection{The lepton system}

The comparatively long lifetime and clean ionization profile of muons
make the lepton kinematics in CC reactions particularly amenable to
precise measurement. NOvA reconstructs muons as tracks and separates
them from the hadronic background using a $k$-nearest neighbors (kNN)
algorithm trained with four variables: the track length, the longitudinal
energy profile ($dE/dx$), the scattering along the track, and the
fraction of energy in the neutrino event associated with the track.
The distribution of the resulting classifier is shown in figure \ref{fig:remid};
the observed data distribution is well-described by the prediction.
Events which have $\text{Muon ID}>0.3$ and whose energy is contained
inside a fiducial volume buffered from the edges of the detector by
two cells are retained as candidate $\nu_{\mu}$ CC events. The predicted
resolutions in both muon energy and angle for this sample are very
good (averaged over the sample, $\unit[50]{MeV}\rightarrow3.8\%$
and $4^{\circ}\rightarrow1.6\%$, respectively), as indicated in figure
\ref{fig:Resolutions}. A doubly-differential cross section measurement
in these variables is currently in progress; the influence of systematic
effects (such as energy scales and the flux prediction) is currently
under investigation.
\begin{center}
\begin{figure}
\centering{}\includegraphics[width=0.75\textwidth]{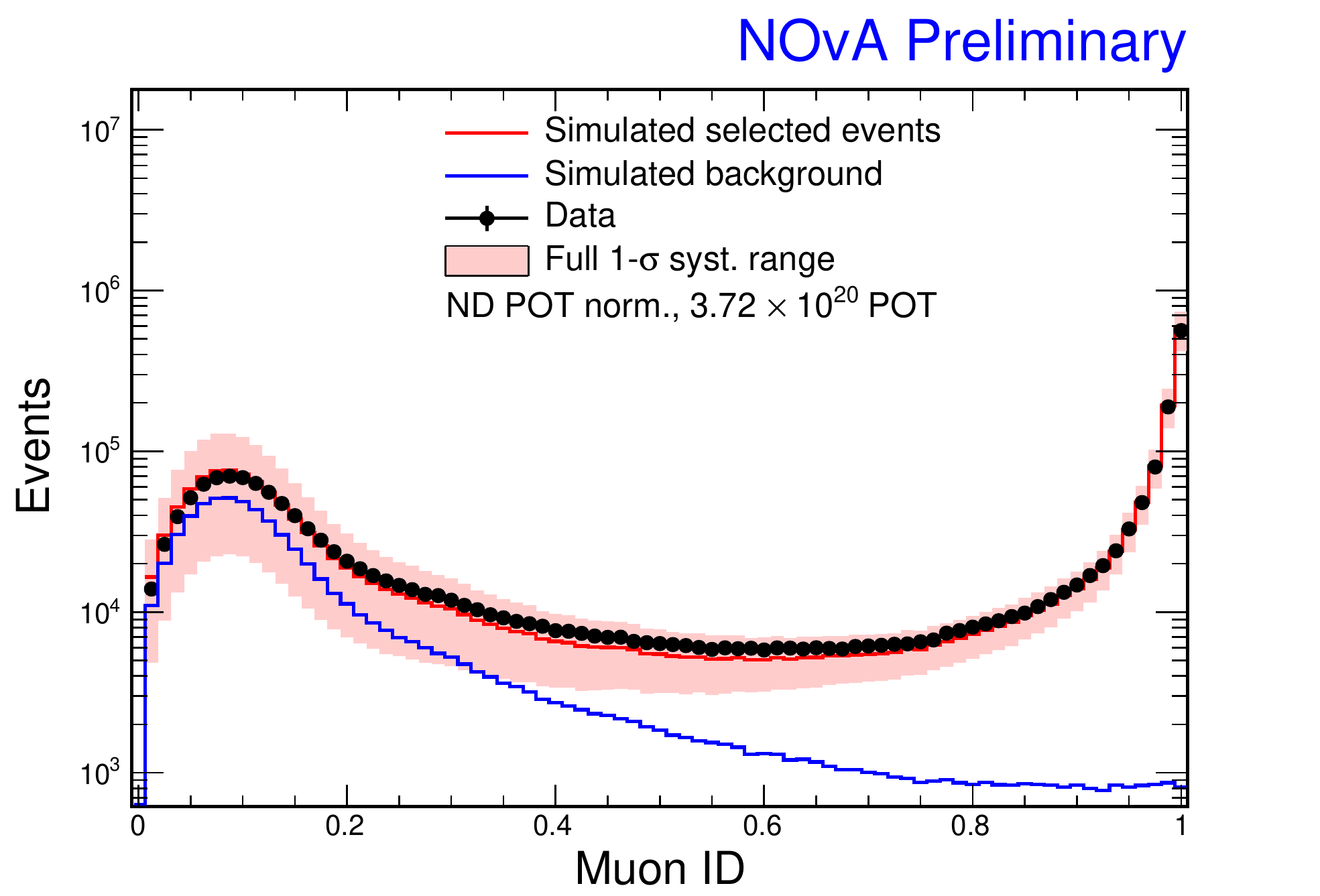}\caption{\label{fig:remid}Predicted distribution of muon particle identification
classifier described in the text for tracks ($\nu_{\mu}$CC signal,
red line; other predicted reactions, blue line) compared to ND data
(black points). Events with $\text{Muon ID}>0.3$ are retained as
candidate $\nu_{\mu}$ CC events.}
\end{figure}
\par\end{center}

\begin{center}
\begin{figure}
\begin{centering}
\includegraphics[width=0.45\textwidth]{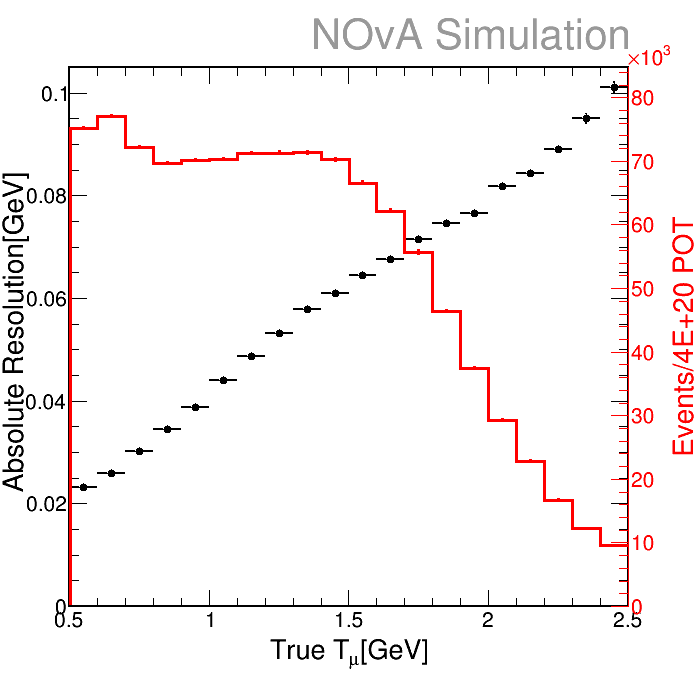}\includegraphics[width=0.45\textwidth]{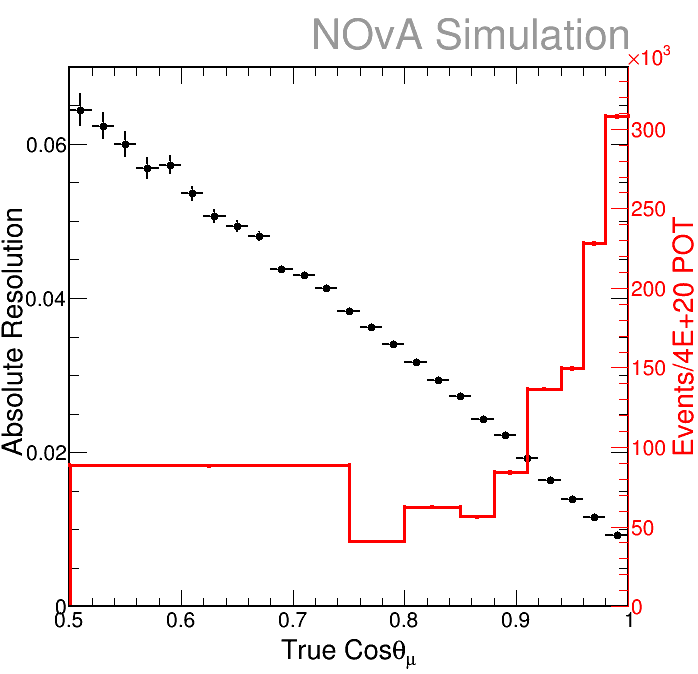}\caption{\label{fig:Resolutions}Predicted resolutions (black dots) compared
to predicted event distributions (red lines) in muon energy (left)
and cosine of the muon angle (right) for selected signal candidates.}
\par\end{centering}
\end{figure}
\par\end{center}

\subsection{The hadronic system}

Because NOvA is a tracking calorimeter, it offers detailed reconstruction
of the hadronic part of $\nu_{\mu}$ CC interactions as well. Here
the effect of the nucleus on neutrino interactions takes center stage;
we observe clear evidence for an extra reaction type beyond those
predicted by default GENIE 2.10.4 lying in between the quasielastic
(QE) and baryon resonance (RES) channels in momentum transfer variables
(where $E_{\mu}$ and $E_{had}$ are the reconstructed muon and non-muon
energies in the system):
\begin{equation}
\begin{gathered}q_{0}=E_{had}\\
E_{\nu}=E_{\mu}+E_{had}\\
Q^{2}=2E_{\nu}\left(E_{\mu}-p_{\mu}\cos(\theta_{\mu})-M_{\mu}^{2}\right)\\
\left|\vec{{q}}\,\right|=Q^{2}+q_{0}
\end{gathered}
\end{equation}

\begin{figure}
\begin{centering}
\includegraphics[width=0.75\textwidth]{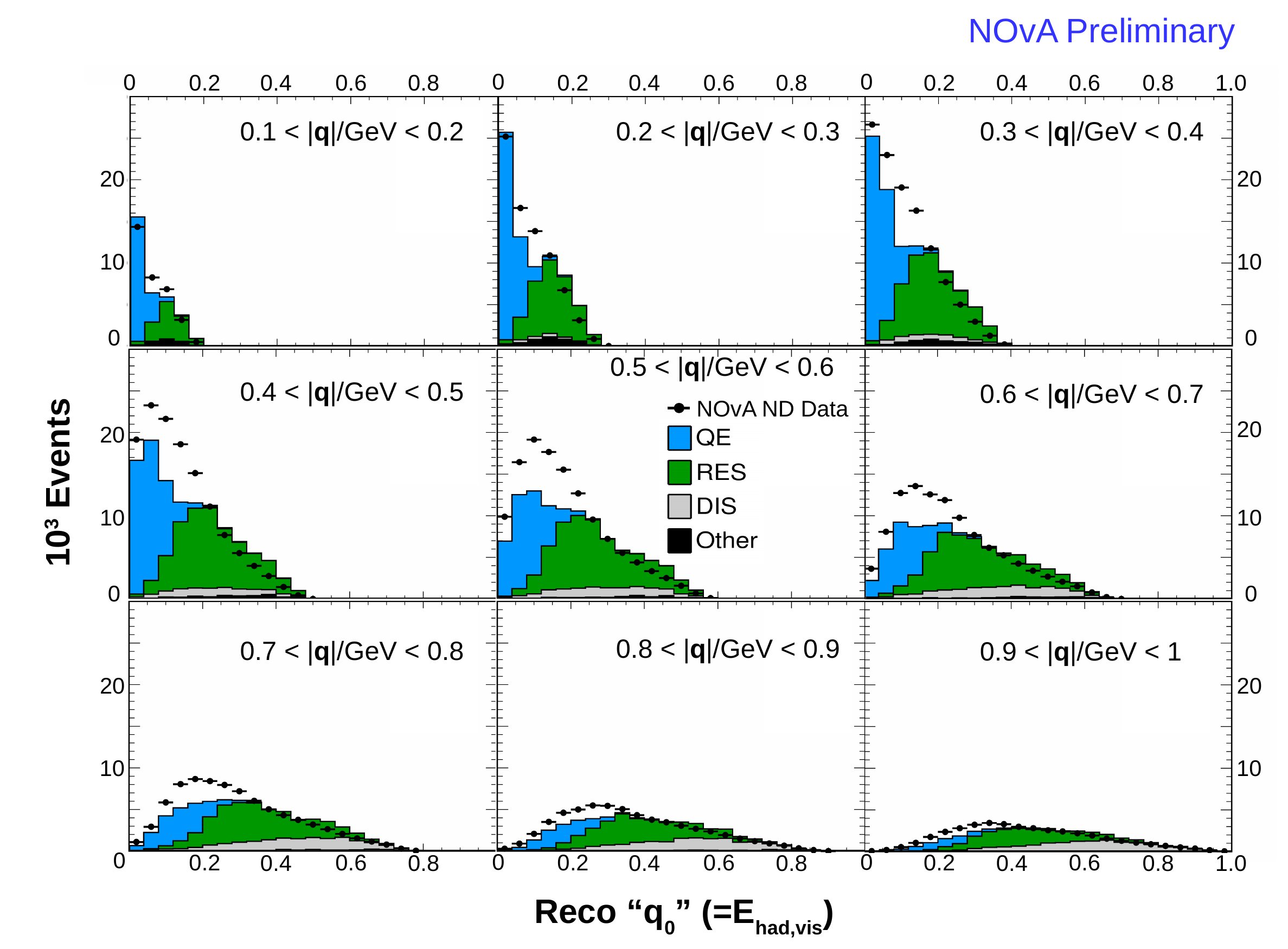}
\par\end{centering}
\centering{}\caption{\label{fig:q0q3 FA}Simulated reactions from GENIE 2.10.4 broken down
by true reaction type (colored areas) compared to NOvA ND data (black
points) in reconstructed energy transfer $q_{0}$, divided into slices
of momentum transfer $|\vec{q}|$ (both described in the text).}
\end{figure}

This is illustrated in figure \ref{fig:q0q3 FA}. Inspired by recent
work in neutrino scattering\cite{minerva-q0q3}, we interpret this
absence as the lack of a model for a two-particle, two-hole (2p2h)
process, where the neutrino scatters from a nucleus and ejects two
of the nucleons (which were previously in some kind of correlated
state) together.

\begin{figure}
\begin{centering}
\subfloat[\label{fig:Ehad FA}No MEC]{\begin{centering}
\includegraphics[width=0.48\textwidth]{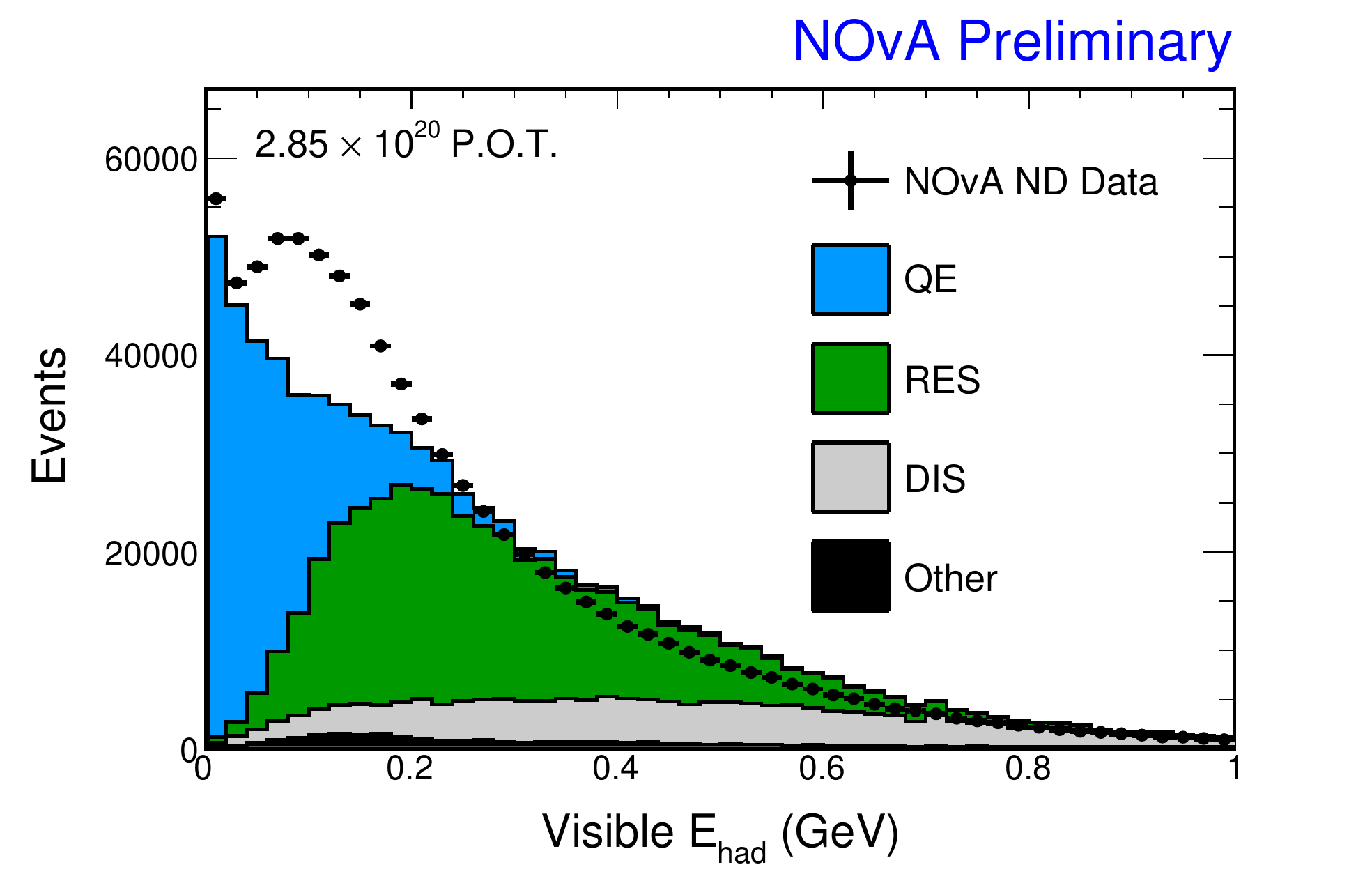}
\par\end{centering}
}\subfloat[\label{fig:Ehad Tufts}Tuned Empirical MEC]{\begin{centering}
\includegraphics[width=0.48\textwidth]{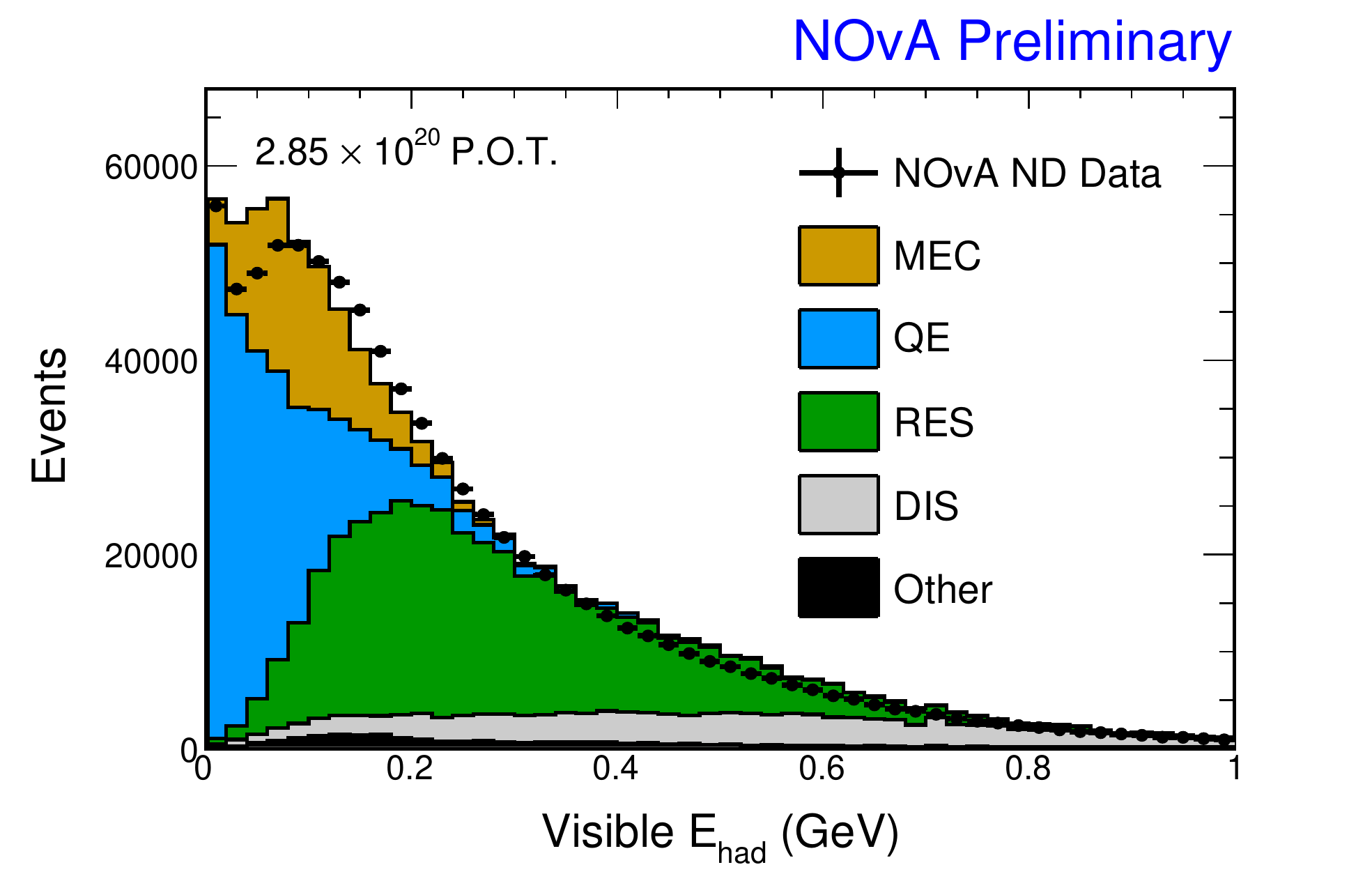}
\par\end{centering}
}\caption{Visible hadronic energy distributions in ND selected $\nu_{\mu}$
CC events before (left) and after (right) the addition of GENIE 2.10.4
``Empirical MEC'' constrained as discussed in the text.}
\par\end{centering}
\centering{}
\end{figure}

GENIE 2.10.4 does ship with an ``optional'' (not enabled by default),
mostly empirical model for 2p2h reactions\cite{katori-empirical-mec},
``Empirical MEC \footnote{Meson Exchange Currents (MEC) are one predicted class of 2p2h which
have generated intense theoretical interest in recent years. Good
summaries of the various strategies can be found elsewhere.\cite{summary-2p2h}}'' (previously called ``Dytman MEC,'' after its author). Because
it is unclear whether the kinematic assumptions built in to this model
that were constructed largely from observations at lower $E_{\nu}$
should extrapolate correctly to NOvA's neutrino energy range, we further
modify this model as follows:
\begin{enumerate}
\item We reverse the linear turn-off of the cross-section between 1 and
\unit[5]{GeV} (so that the Empirical MEC cross section becomes a
constant fraction of the QE one) since there are recent indications
that 2p2h exists with similar strength at energies above \unit[5]{GeV}.\cite{minerva-q0q3}
\item We reverse the fraction of scattering from neutron-neutron and neutron-proton
pairs in the model to 20\% and 80\%, respectively, based on indications
from electron scattering\cite{elec-scat-frac} and expectations from
theoretical expectations in neutrino scattering\cite{superscaling-nn-np}.
\footnote{The typo that led to the need for this correction has been corrected
in GENIE 2.12.}
\item We apply a momentum-transfer-dependent weight derived from our ND
data as described in the next paragraph.
\end{enumerate}
To construct weights that constrain the Empirical MEC to better fit
our observed data, we first examine the data excess in $|\vec{q}|$
(effectively the difference of the integrals of data and simulation\footnote{After applying the correction to non-resonant $1\pi$ production from
neutrons suggested by Rodrigues \textit{et al.}\cite{rod-genie-tune}} in each panel of figure \ref{fig:q0q3 FA}). We reweight the Empirical
MEC such that it agrees with the data excess in this variable. To
set the fourth component of the four-momentum transfer, $q_{0},$
we fix it to the shape of the predicted $q_{0}$ distribution in each
bin of $|\vec{q}|$ taken from the GENIE quasielastic channel. This
somewhat underestimates the $E_{had}$ in the observed distribution,
as illustrated in figure \ref{fig:Ehad Tufts}, but the overall agreement
relative to the untuned version (figure \ref{fig:Ehad FA}) is substantially
improved. The GENIE 2.10.4 prediction with tuned Empirical MEC is
the base prediction for current oscillation analysis efforts, including
those discussed elsewhere in this volume.

\section{$\nu_{e}$ charged-current inclusive scattering}

Electron neutrinos are expected to undergo the same types of reactions
and their interactions are expected to experience the same types of
nuclear effects as $\nu_{\mu}$, up to the influence of the difference
in the charged lepton masses. Understanding whether this is actually
the case is very important for oscillation experiments like NOvA,
for which the interactions of $\nu_{e}$ appearing via oscillation
from a $\nu_{\mu}$ beam comprise a critical signal channel. However,
at energies around several GeV, until recently it has been challenging
to accumulate enough $\nu_{e}$ interactions to make statistically
significant measurements. The very intense NuMI beam used by NOvA,
on the other hand, has about a 1\% admixture of $\nu_{e}$, opening
the door for high-statistics investigation.

\begin{figure}
\centering{}\subfloat[\label{fig:LID}]{\includegraphics[width=0.48\textwidth]{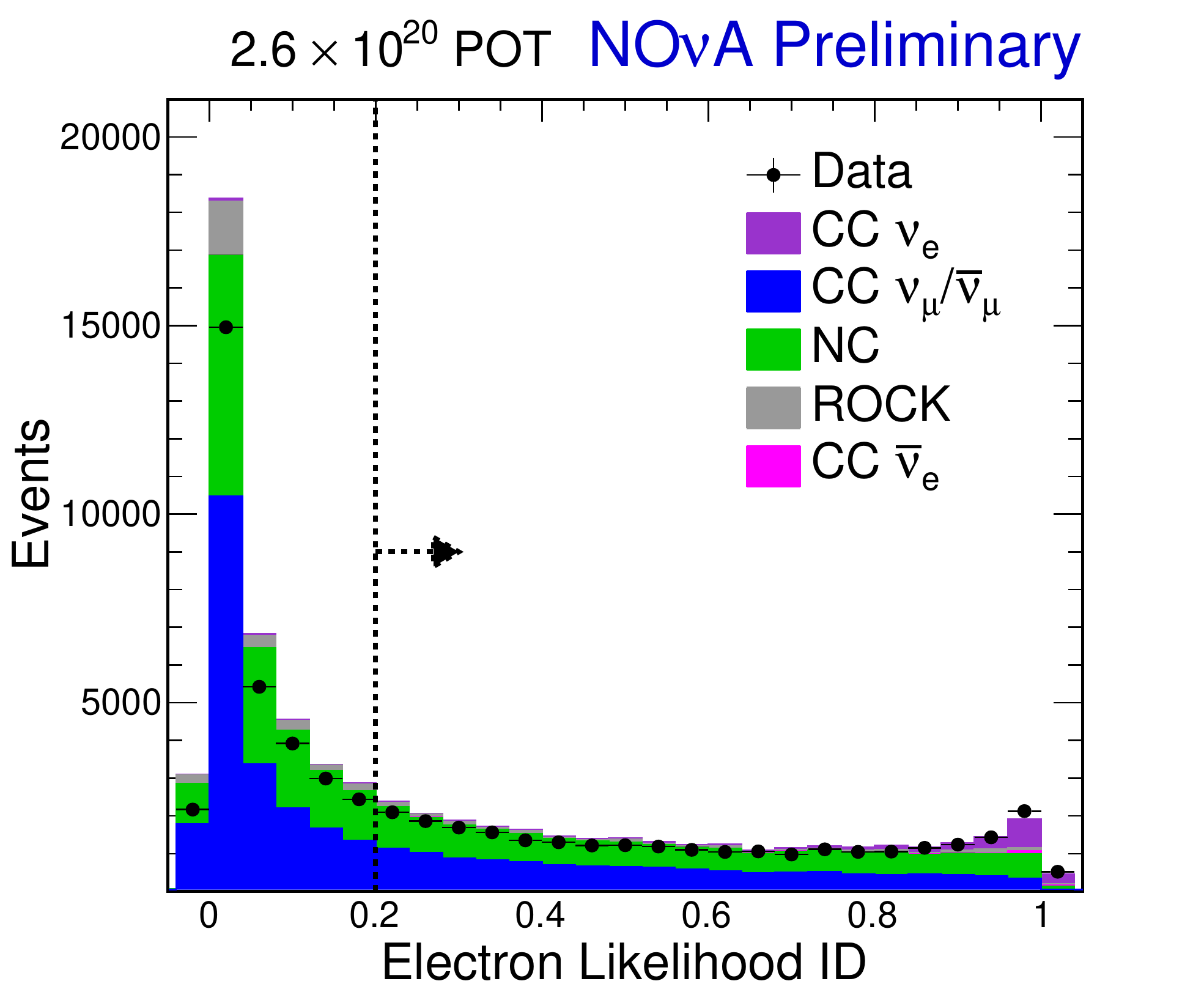}

}\subfloat[\label{fig:electron BDT}]{\includegraphics[width=0.48\textwidth]{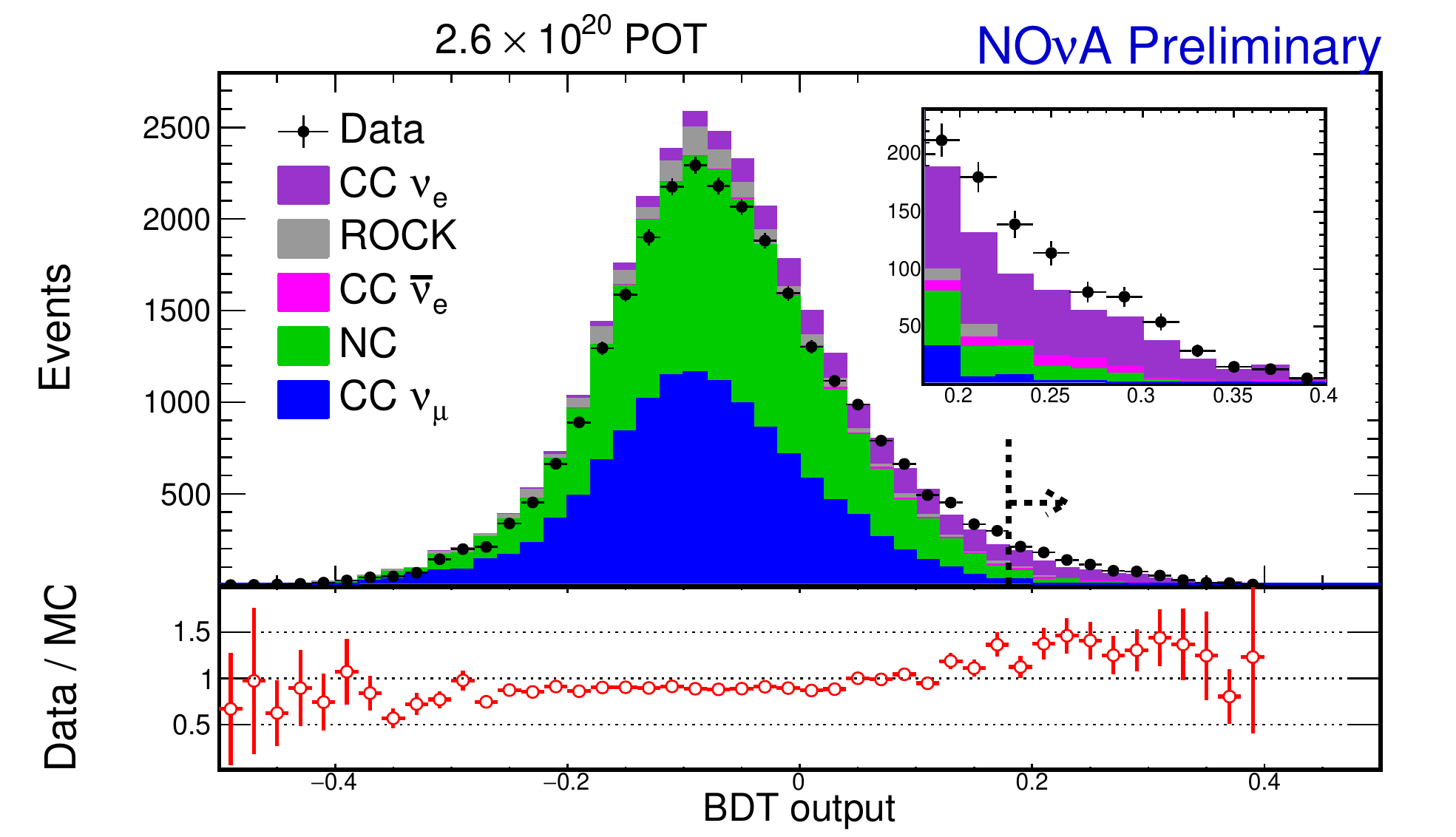}

}\caption{Performance of variables used to select $\nu_{e}$ CC interactions,
as described in the text: likehood classifer preselector (left); final
boosted decision tree output (right). The inset in the right plot
is a zoom showing only the distribution above the cut.}
\end{figure}

For a cross section analysis, NOvA begins selecting $\nu_{e}$ interactions
using a likelihood classifier constructed from the longitudinal energy
profiles of various particle templates; the performance of this classifier
(after a baseline selection requiring containment and rejecting especially
minimum-ionizing tracks to reject $\nu_{\mu}$ CC), and the selection
cut made on it, is illustrated in figure \ref{fig:LID}. Once this
electromagnetic cascade-enhanced sample is obtained, further purification
is accomplished using a boosted decision tree using shower shape variables
(both longitudinal and transverse); its performance is shown in fig.
\ref{fig:electron BDT}. Studies of sideband regions in these variables
are underway in order to better constrain the predicted backgrounds
and understand what the dominant uncertainties will be for a cross
section.

\section{Constraining neutrino flux with $\nu-e$ elastic scattering}

The neutrino flux prediction is an essential ingredient to any cross
section measurement because it represents the normalization coefficient
as a function of neutrino energy; traditionally flux uncertainties
comprise the largest source of error for extracted cross sections.
This owes primarily to the fact that \textit{ab initio} calculations
of horn-focused neutrino beams like NuMI depend on predictions of
the strong-force dynamics of protons colliding (and re-interacting)
with complex molecular targets like graphite, which are difficult.
However, it is in principle possible to constrain the flux prediction
using an \textit{in situ} measurement of a neutrino scattering process
with a well-understood cross section. Because of the complexities
of neutrino interactions with nuclei, however, purely leptonic processes
like $\nu+e\rightarrow\nu+e$ scattering (neutrinos with atomic electrons)
are the the reactions most amenable to use in this fashion. Unfortunately,
the cross section of $\nu+e\rightarrow\nu+e$ scattering is suppressed
relative to nucleon scattering by the ratio of the electron to nucleon
masses and other kinematic factors, resulting in $\sigma_{\nu-e}/\sigma_{\nu-N}\sim10^{-4}$.
Therefore statistics are typically low in this channel.

\begin{figure}
\centering{}\subfloat[\label{fig:etheta2}]{
\begin{centering}
\includegraphics[width=0.48\textwidth]{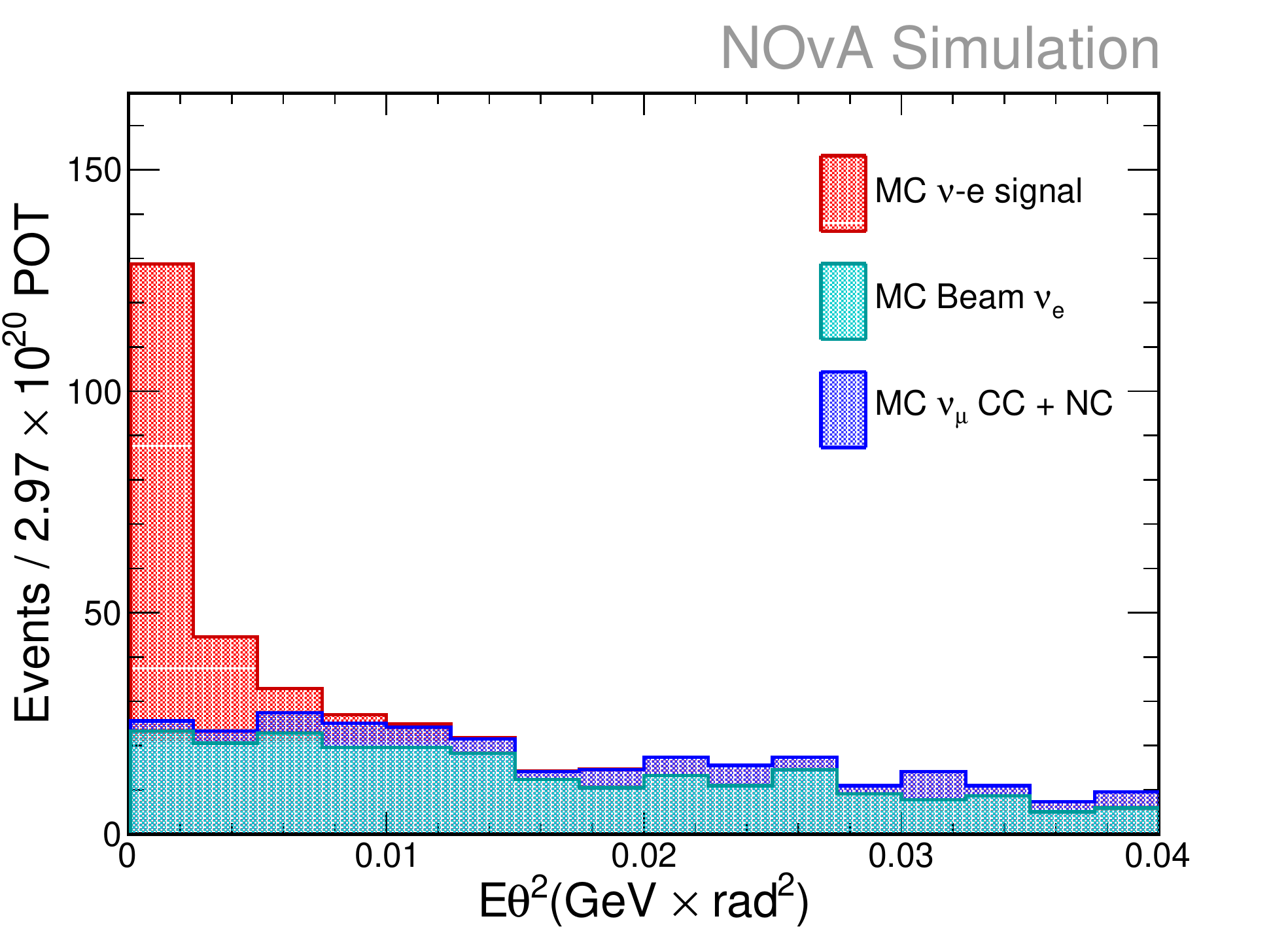}
\par\end{centering}
}\subfloat[\label{fig:Eelecton}]{\begin{centering}
\includegraphics[width=0.48\textwidth]{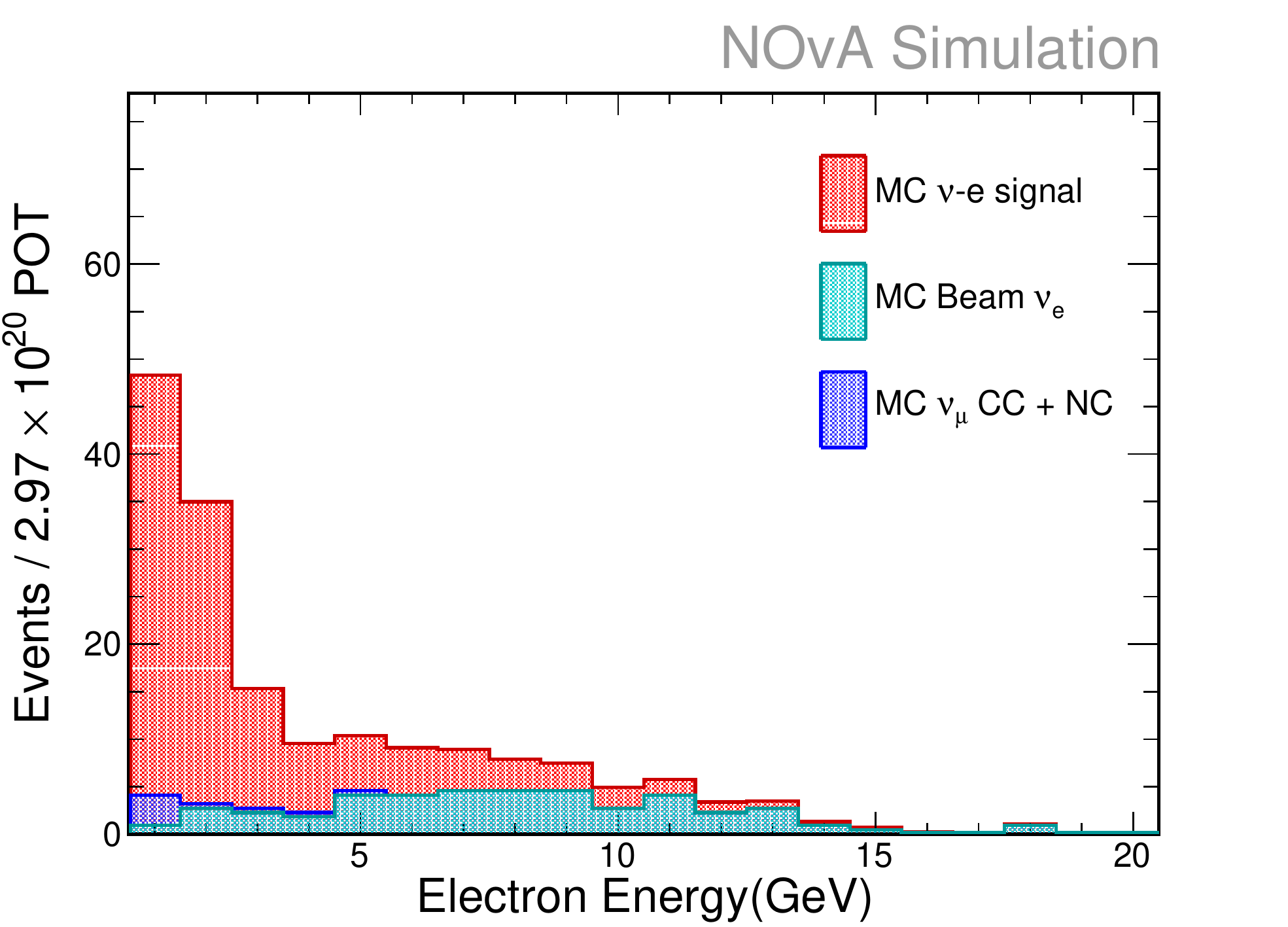}
\par\end{centering}
\begin{centering}
\par\end{centering}
\centering{}}\caption{Electron $E\theta^{2}$ (left), where events with $E\theta^{2} < \unit[0.005]{GeV\times rad^{2}}$
are considered signal candidates, and resulting electron energy distribution
(right) in the $\nu+e\rightarrow\nu+e$ scattering analysis.}
\end{figure}

As in the $\nu_{e}$ CC case, NOvA uses two PID classifiers to identify
candidate electron showers for this analysis: one that distinguishes
between electromagnetic showers and other backgrounds, and one that
specifically distinguishes between electron-induced and photon- or
neutral pion-induced showers. After selections on these variables,
we employ a cut at $\unit[0.005]{GeV\times rad^{2}}$ on the kinematic
variable $E_{e}\theta_{e}^{2}$, which is limited to very small values
by the kinematics of the interaction itself, to further enrich the
signal; this is illustrated in figure \ref{fig:etheta2}. The resulting
electron energy spectrum, which will be used to constrain the flux,
is shown in figure \ref{fig:Eelecton}. Currently efforts are being
devoted to quantifying the size of uncertainty in the signal efficiency
and background cross section and flux predictions. It is expected
that this technique will constrain the flux normalization to around
10\% uncertainty.

\section*{Acknowledgments}

NOvA is supported by the US Department of Energy; the US National
Science Foundation; the Department of Science and Technology, India;
the European Research Council; the MSMT CR, Czech Republic; the RAS,
RMES, and RFBR, Russia; CNPq and FAPEG, Brazil; and the State and
University of Minnesota. We are grateful for the contributions of
the staffs of the University of Minnesota module assembly facility
and NOvA FD Laboratory, Argonne National Laboratory, and Fermilab.
Fermilab is operated by Fermi Research Alliance, LLC under Contract
No. De-AC02-07CH11359 with the US DOE.


\begin{thebibliography}{1}
\bibitem{rod-genie-tune}P. Rodrigues \textit{et al.}, \Journal{Eur. Phys. J.}{C76}{474}{2016}.

\bibitem{minerva-q0q3}P. Rodrigues \textit{et al.} (MINERvA Collaboration),
\Journal{Phys. Rev. Lett.}{116}{071802}{2016}.

\bibitem{katori-empirical-mec}T. Katori, \Journal{AIP Conf. Proc.}{1663}{030001}{2015}.

\bibitem{summary-2p2h}A. Ankowski \textit{et al.}, \Journal{Phys. Rev.}{D93}{113004}{2016}.

\bibitem{elec-scat-frac}R. Subedi \textit{et al.}, \Journal{Science}{320}{1476}{2008}.

\bibitem{superscaling-nn-np}I. Ruiz-Simo\textit{ et al.}, \Journal{Phys. Lett.}{B762}{124}{2016}.
\end{thebibliography}
\end{document}